\documentclass{article}
\usepackage{latexsym}
\usepackage{bm}
\usepackage{amssymb}
\usepackage{latexsym}
\usepackage{graphicx}
\usepackage{layout}
\usepackage{amsmath}
\usepackage{textcomp}
\usepackage{setspace}
\usepackage{lineno}

\begin{document}

\title{Induced spin velocity of the Earth and its influence to the seasonal variation of the Earth's angular velocity}

\author{Kostadin Tren\v{c}evski$^1$, Emilija Celakoska$^2$\\
     $^1$Faculty of Natural Sci. and Math., Ss. Cyril and Methodius University, \\
     P.O.Box 162, Skopje, Macedonia, e-mail: kostadin.trencevski@gmail.com\\
     $^2$Faculty of Mechanical Engineering, Ss. Cyril and Methodius University,\\
     Karpos II bb., Skopje, N. Macedonia, e-mail: emilija.celakoska@mf.edu.mk
}

\date{}

\maketitle



\begin{abstract}
We examine the induced spin velocity in case of the Earth. Spin velocity is induced from the conversion 
of a constrained spatial rotation into a spatial displacement. Its effects on Earth as a celestial body are 
consequences of its properties and they are examined in detail. The induced spin velocity has influence 
to the semiannual variation of the length of day. The annual and semiannual variation of the length of 
day are considered separately. The measured value in case of the semiannual variation of the length 
of the day is 5.44\% more than the predicted, while the measured value in case of the 
annual variation of the length of the day is 5.36\% less than the predicted.


\end{abstract}

\section{Introduction}

The idea of multidimensional manifold $(n>4)$ describing the space-time and the corresponding 
geometrical quantities is old more than one century, with some contributions of Einstein, after the formulation of the general relativity theory.
Particularly, multidimensional geometry of time with different points of view are 
proposed or analyzed in recent times \cite{11,1,2,3,4,6,7,8,9}. This paper is a continuation of the papers \cite{MB,VB,Tensor2009,CEJP2013,DGDS2013},
and improvement of \cite{CEJP2013} on that subject. Gravitation in the corresponding multidimensional space-time is recently published in \cite{TC}.

We denote by $x$, $y$ and $z$ the coordinates in
${\mathbb R}^3$ and parameterize the bundle of all moving orthonormal frames
by the nine coordinates
$x,y,z,$ $x_s,y_s,z_s,$ $x_t,y_t,z_t,$
where the first six of them
parameterize the subbundle with the fiber $SO(3,{\mathbb R})$.
So, we call it 3+3+3-dimensional model \cite{MB,VB,Tensor2009,CEJP2013,DGDS2013},
since we relate to each body 3 coordinates for the position, 3 coordinates for the spatial rotation
and 3 coordinates for the velocity.
The 3+3+3-model is built on three three-dimensional sets:
space ($S$) which is homeomorphic to $S^3$,
spatial rotations ($SR$) which is also homeomorphic to $S^3$ and
velocity ($V$). 

\section{Some preliminaries of spinning bodies in a gravitational field}

We give some preliminaries according to \cite{Filomat} and also, some improvements.
It is known that the Lorentz group $O_{+}^{\uparrow}(1,3)$ is isomorphic to $SO(3,{\mathbb C})$,
and both of them are homeomorphic to $SO(3,{\mathbb R})\times {\mathbb R}^3$.
Instead of these real $6\times 6$-matrices
we are interested for the product $S\times SR$,
which can be considered as a fiber and Lie group $G$ of a principal bundle over the base $V$.
So, we consider this group for a fixed inertial coordinate system up to a translation and spatial rotation
and the coefficient $\sqrt{1-\frac{v^2}{c^2}}$ will not have any role.
This group is analogous to the
group of all rotations and translations in the 3-dimensional Euclidean space.
The Lie algebra of $G$ is given by
\begin{equation}
\left [\begin{array}{ccc}C & &B \\ B& & C\end{array}\right ].\label{12.1}
\end{equation}
where $B$ and $C$ are antisymmetric $3\times 3$ matrices.

The group $G$ is isomorphic to the group $Spin(4)$ \cite{DGDS2013}.
While the Lorentz group reduces to the group of Galilean transformations when the velocities are small,
the transformations of the group $G$ reduce to the affine group
of all rotations and translations in the Euclidean space in case of short translations, i.e. matrices of type
$\left [\begin{array}{ccc}M & &\vec{h}^T\\0& & 1\end{array}\right ]$,
where $M\in SO(3,{\mathbb R})$ and ${\vec{h}}^T$ is the vector of translation.

If a rigid body is spinning, there may appear a constraint for
the spatial rotation, because there is no freedom of a chosen point to rotate according to its own trajectory.
As a consequence, there may appear a displacement which is called spin displacement, because it appears in case of spinning bodies.
The property of conversion from an constrained spatial rotation into a spatial displacement is a basic property of the space.
This displacement induces the so-called
{\em induced spin velocity} or simply {\em spin velocity} \cite{CEJP2013} and will be denoted by large letter $V$.

The spin motion (displacement) has the following properties.

i)
The spin velocity is non-inertial,
because it can be conceived just like a displacement in the space.

ii) Instead the Lorentz transformations for these velocities, we have analogous transformations where the coefficient
$\sqrt{1-\frac{V^2}{c^2}}$ does not appear.

iii) If the spin velocity of
any point is constrained completely or partially, then the constrained part converts into inertial velocity with opposite sign.

Let us consider a trajectory over a spinning sphere, which rests in our
coordinate system, but it is under the gravitational acceleration or any mechanical force.
We assume that the barycenter is at the coordinate origin, that
at the initial moment the spin axis is determined by $\vec{b}^*=(0,0,1)$ and at the initial moment the considered point has coordinates
$(r\cos\alpha ,r\sin\alpha ,h)$.

In order to calculate the spin velocity we will use the group of affine transformations in three-dimensional Euclidean space.
Its Lie algebra has the following form
\begin{equation}
A=\left [\begin{array}{cccc}0 & -\varphi_z & \varphi_y & s_x\\
\varphi_z & 0 & -\varphi_x & s_y\\
-\varphi_y & \varphi_x & 0 & s_z\\
0 & 0 & 0 & 0 \end{array}\right ], \label{l.a.}
\end{equation}
where $\vec{\varphi}=(w_x,w_y,w_z)t$, $\vec{w}$ is the angular velocity of the sphere,
$t$ is short time and
$\vec{s}=(g_x,g_y,g_z)t^2/2$ is
small translation as a consequence of the acceleration $\vec{g}$. The acceleration $\vec{g}$ represents mainly the
gravitational acceleration, but also the acceleration which keeps the spinning body to avoid free fall motion.
In case of free fall motion of the spinning body,
we should consider $\vec{g}$ only as vector of gravitational acceleration. But, otherwise, 
we should additionally take into account the acceleration of the horizontal plane, 
which acts to the spinning body in the direction of the spinning axis. Indeed, in this case, 
the total acceleration is $\vec{g}-\vec{b}^*[\vec{g}\cdot \vec{b}^*]$. In a special case,
when $\vec{b}^*$ is collinear to $\vec{g}$, then the total acceleration is 0.

The quantities $\vec{\varphi}$ and $\vec{s}$ may depend on time, so we use the Taylor series. Since
$\vec{\varphi}(0)=0$, $\vec{s}(0)=0$, and $\vec{s}'(0)=0$ we obtain

$$\vec{\varphi}(t)=\vec{\varphi}(0)+\vec{\varphi}'(0)\frac{t}{1!}+\vec{\varphi}''(0)\frac{t^2}{2!}+\cdots
=\vec{w}t+\vec{w}'\frac{t^2}{2!}+\vec{w}''\frac{t^3}{3!}+\cdots $$
and

$$\vec{s}(t)=\vec{s}(0)+\vec{s}'(0)\frac{t}{1!}+\vec{s}''(0)\frac{t^2}{2!}+\cdots
=\vec{g}\frac{t^2}{2}+\vec{g}'\frac{t^3}{6}+\cdots . $$
After these replacements into (\ref{l.a.}) the required trajectory is determined by the matrix
$I+A+\frac{A^2}{2!}+\frac{A^3}{3!}+\cdots$. Then the image $(x(t),y(t),z(t))$ of the starting vector
$(r\cos\alpha ,r\sin \alpha ,h)$, where $\alpha$ is an arbitrary parameter of the circle, is given by the equality
\begin{equation}
\left [\begin{array}{c}x(t)\\ y(t)\\ z(t)\\ 1\end{array}\right ]
=(I+A+\frac{A^2}{2!}+\frac{A^3}{3!}+\cdots )
\left [\begin{array}{c}r\cos \alpha \\ r\sin \alpha \\ h \\ 1\end{array}\right ].\label{curve}
\end{equation}
Hence $\vec{r}=(x(t),y(t),z(t))$ is well defined, and the first three derivatives are
\begin{equation}
\vec{r}'=(-w\sin \alpha ,w\cos \alpha ,0)r,\label{r'}
\end{equation}
$$\vec{r}''=
(-rw^2\cos \alpha -rw'_z\sin \alpha +hw'_y+g_x,
-rw^2\sin \alpha +rw'_z\cos \alpha -hw'_x+g_y,$$
\begin{equation}
-rw'_y\cos \alpha +rw'_x\sin \alpha +g_z),\label{r''}
\end{equation}
\begin{eqnarray}
\vec{r}'''=
(-3rww'_z\cos \alpha +rw^3\sin \alpha -w''_zr\sin \alpha +hw''_y+\frac{3}{2}hww'_x,\nonumber    \\
-3rww'_z\sin \alpha -rw^3\cos \alpha +w''_zr\cos \alpha -hw''_x+\frac{3}{2}hww'_y,\nonumber \\
(-w''_y+\frac{3}{2}ww'_x)r\cos \alpha + (w''_x+\frac{3}{2}ww'_y)r\sin \alpha )-
\frac{3}{2}(\vec{g}\times \vec{w})+\vec{g}'.\label{r'''}
\end{eqnarray}

Any point of the spinning sphere intends to move in its own osculating plane, orthogonal to the binormal vector
$\vec{b}$, but as a part of the sphere at the chosen moment all points will move in the plane which is orthogonal to
the vector $\vec{b}^*$. Note that in general case $\vec{b}\neq \vec{b}^*$. We will assume further that $\frac{d\vec{b}^*}{dt}<<w$.

If there are no constraints, the Frenet antisymmetric matrix
\begin{equation}
\left [\begin{array}{ccc}0 & k & 0\\ -k & 0 & \tau\\ 0 & -\tau & 0\end{array}\right ]ds
\end{equation}
corresponds to angular rotation of the trihedron $(\vec{t},\vec{n},\vec{b})$ by (\cite{BP}, sec.28)
\begin{equation}
\tau \vec{t}ds +k \vec{b}ds.\label{natural}
\end{equation}
One can explain why an
arbitrary point of the considered trajectories over the sphere $r^2+h^2=const$ tends to rotate with accordance to the rotation of
the trihedron $(\vec{t},\vec{n},\vec{b})$, but we will omit this discussion.

Assume that the considered point, which moves on the considered trajectory, may be displaced without constraint.
There may exist different approaches in determining the spin velocity, but all of them have the same approximation when
$\tau <<k$. In \cite{CEJP2013} is given one such procedure. Analogously to the invariant 
$dx^2+dy^2+dz^2-c^2dt^2$ in the Special Relativity, in the space $SR\times S$ there exist two invariants
\begin{equation}
I_1=(d\vec{\eta })^2+(d\vec{\xi})^2, \qquad I_2=d\vec{\eta }\cdot d\vec{\xi},\label{I1I2}
\end{equation}
where $d\vec{\eta }$ is vector of spatial displacement caused by the space (i.e. translation), while $d\vec{\xi}$ is displacement
caused by the rotation given by (\ref{natural}). The property that they are invariant means that they remain unchanged
independently of (non)existence of constraints.  Now, we have the following theorem
(\cite{Filomat}) presented in Fig. 1.

{\bf Theorem.} {\em
The induced spin velocity of arbitrary point on a spinning sphere, whose center rests in our coordinate system at the initial moment,
is given by
\begin{equation}
\vec{V}=-\frac{\tau k}{k^2+\tau^2}rw\vec{b}-\frac{\tau^2}{k^2+\tau^2}rw\vec{t},\label{sv}
\end{equation}
where $w$ is the angular velocity and $r$ is distance to the axis of the sphere.}

\begin{figure}
\begin{center}
 \includegraphics[scale=1.0]{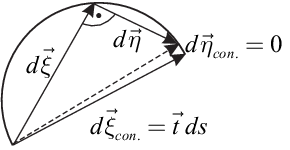}
 \label{fig:0}
 \caption{Spin velocity on a spinning sphere. $d\vec{\eta}$ denotes spatial displacement, while $\vec{\xi}$ denotes
displacement on a circular motion caused by rotation, while "con" denotes displacement with constraints. }
\end{center}
\end{figure}

{\it Proof.} The vector of displacement caused by the rotation for angle (\ref{natural}) is orthogonal with the unit vector $-\vec{n}$
and the vector (\ref{natural}) and hence,

$$d\vec{\xi} =\mu [(\tau \vec{t} +k \vec{b})\times (-\vec{n})ds]=\mu (-\tau \vec{b}+k\vec{t})ds.$$

The vector of displacement caused by rotation (\ref{natural}), where $\tau \approx 0$ since $d\vec{b^*}/dt<<w$ is
$d\vec{\xi}_{con}=\vec{t}ds$, while $d\vec{\eta}_{con}=0$ because the center of the spinning sphere rests.
According to (\ref{I1I2}) we have the following system

$$(d\vec{\eta})^2+(\mu (-\tau \vec{b}+k\vec{t})ds)^2=(\vec{t}ds)^2,$$

$$(d\vec{\eta})\cdot(\mu (-\tau \vec{b}+k\vec{t})ds)= 0.$$
Using also that

$$(\mu (-\tau \vec{b}+k\vec{t})ds)+(\vec{\eta}ds) = \vec{t}ds,$$
one can easily obtain that $\mu =k/(k^2+\tau^2)$. The spin velocity $\vec{V}$ is indeed the vector
$\frac{d\vec{\xi}}{dt}-\frac{d\vec{\xi}_{con}}{dt}= \frac{d\vec{\eta}_{con}}{dt}-\frac{d\vec{\eta}}{dt}$
and now using that $ds=rwdt$, it is given by (\ref{sv}).

According to (\ref{sv}) we notice that

$i)$ $\vert \vec{V}\vert =\vert\frac{\tau}{\sqrt{k^2+\tau^2}}rw\vert \leq \vert rw\vert $,

$ii)$ $\vert\vec{b}\cdot \vec{V}\vert \leq \vert rw\vert /2$,
and

$iii)$ $\vec{V}$ is collinear with the vector of rotation (\ref{natural}).
If $\lambda =\tau/ k$, then the spin velocity becomes
\begin{equation}
\vec{V}=-\frac{\lambda}{1+\lambda^2}rw\vec{b}-\frac{\lambda^2}{1+\lambda^2}rw\vec{t}.\label{sv1}
\end{equation}
We will use the notations $\vec{V}_b=-\frac{\lambda}{1+\lambda^2}rw\vec{b}$ and
$\vec{ V}_t=-\frac{\lambda^2}{1+\lambda^2}rw\vec{t}$.

The spin velocity $\vec{V}$ can be decomposed as
\begin{equation}
\vec{V}=\frac{(\vec{V}\cdot \vec{r})\vec{r}}{r^2}+[\vec{V}-\frac{(\vec{V}\cdot \vec{r})\vec{r}}{r^2}],\label{decom}
\end{equation}
where $\vec{r}$ is the radius vector starting from the barycenter of the spinning body. The first component from the right side
in (\ref{decom}),
i.e. the radial component, participates in the global spin motion (or displacement), which we call a global spin velocity.
So, the global spin velocity is given as the following sum over large number of small particles of the spinning body
\begin{equation}
\vec{V}=\sum_i \frac{m_i(\vec{V}_i\cdot \vec{r}_i)\vec{r}_i}{Mr_i^2},\label{gsv}
\end{equation}
where $M$ is the mass of the body. The summation in this formula in general case reduces to 
averaging with respect to the angular parameter $\alpha \in [-\pi ,\pi ]$.
Since the tangent vector $\vec{t}$ is orthogonal to the radial vector $\vec{r}$, we notice that $\vec{V}_t$ has no influence in
the global spin velocity.

\section{Applications of the spin velocities in case of spinning bodies in the Earth's gravitational field}

Let us consider a spinning circle as a gyroscope, where $w$ is constant and the vector $\vec{b}^*$ rotates with
a constant angular velocity $\Omega$ around the vertical axis, i.e. around the vector $\vec{g}$, so that the
angle $\varphi$ between the vector $\vec{b}^*$ and the vertical axis is constant.
We should distinguish two different cases:

(a) Assume that the gyroscope is in free fall motion, and the equations of motion
in the horizontal plane are
the equations of the projection on the horizontal plane;

(b) Assume that the gyroscope moves on a fixed horizontal plane.

In both cases we can write

$$\vec{b}^*=(a\cos \Omega t,a\sin \Omega t, c),$$
where $a=\sin \varphi$ and $c=\cos \varphi$. Using that $\vec{w}=w\vec{b}^*=w(a\cos \Omega t,a\sin \Omega t, c)$, where
$w$ is constant, we obtain

$$\vec{w}'=\Omega wa(-\sin \Omega t,\cos \Omega t,0),\quad
\vec{w}''=-\Omega^2wa(\cos \Omega t,\sin \Omega t,0).$$
The two different cases differ in application of the gravitational acceleration, as it was discussed in section 2.

(a) The previous formulas should be replaced in the general formulas for arbitrary $\vec{b}^*$,

$$\vec{r}'=wr\vec{t},\qquad
\vec{r}''=-rw^2(\vec{t}\times \vec{b}^*) + \vec{t}\cdot r(\vec{w}'\cdot \vec{b}^*)-
\vec{b}^*(\vec{w}'\cdot \vec{t})r+\vec{g},$$
$$\vec{r}'''=-[rw^3-r(\vec{w}''\cdot \vec{b}^*)]\vec{t}-3rw(\vec{w}'\cdot \vec{b}^*)(\vec{t}\times \vec{b}^*)+
\vec{b}^*(-(\vec{t}\cdot \vec{w}'')+\frac{3}{2}w[(\vec{t}\times \vec{b}^*)\cdot \vec{w}'])r-$$
$$-\frac{3}{2}w(\vec{g}\times \vec{b}^*)
+\frac{d\vec{g}}{dt}.$$
Note that $\vec{w}'\cdot \vec{b}^*=0$, because $w$ is a constant.
The unit tangent vector $\vec{t}$, which is orthogonal to $\vec{b}^*$ at the initial moment $t=0$
can be parameterized by $\vec{t}=(-c\sin \alpha ,\cos \alpha ,a\sin \alpha )$. We use also that $\vec{g}=(0,0,-g)$,
where $g$ is a constant.

In order to avoid large expressions, we will make the calculations at the moment $t=0$, such that

$$\vec{b}^*=(a,0,c), \quad \vec{w}'=a\Omega w(0,1,0),\quad \vec{w}''=-a\Omega^2w(1,0,0).$$
Hence after all these substitutions, for the derivatives of $\vec{r}$ we obtain
$$\vec{r}'=wr(-c\sin \alpha ,\cos \alpha ,a\sin \alpha ),$$
$$\vec{r}''=-rw^2(c\cos \alpha ,\sin \alpha ,-a\cos \alpha ) -
arw\Omega \cos \alpha (a,0,c) -(0,0,g),$$
$$\vec{r}'''=-(rw^3+ra^2w\Omega ^2)(-c\sin \alpha ,\cos \alpha ,a\sin \alpha )-
(a,0,c)ra\Omega w\sin \alpha (\Omega c-\frac{3}{2}w)+$$
$$+\frac{3}{2}awg(0,1,0).$$
Further, we obtain
 $$\vec{r}'\times \vec{r}'''=
-ar^2w^2\Omega \sin \alpha (\Omega c-\frac{3}{2}w)(c\cos \alpha ,\sin \alpha ,-a\cos \alpha )-
\frac{3}{2}rw^2ag\sin \alpha \vec{b}^*,$$
$$(\vec{r}'\times \vec{r}''')\cdot \vec{r}''=
ar^3w^4\Omega \sin \alpha (\Omega c-\frac{3}{2}w)+\frac{3}{2}a^2r^2\Omega w^3g\cos \alpha \sin \alpha -$$
$$-ga^2r^2w^2\Omega \sin \alpha \cos\alpha (\Omega c-\frac{3}{2}w) +\frac{3}{2}g^2rw^2ac \sin \alpha ,$$
$$(\vec{r}',\vec{r}'',\vec{r}''')=-ar^3w^4\Omega \sin \alpha (\Omega c-\frac{3}{2}w)
+ga^2cr^2w^2\Omega^2 \sin \alpha \cos \alpha -
\frac{3}{2}g^2rw^2ac\sin \alpha -$$
$$-3a^2r^2w^3\Omega g\sin\alpha\cos\alpha , $$
$$\vec{r}'\times \vec{r}''=r^2w^3\vec{b}^*-ar^2w^2\Omega\cos \alpha (c\cos \alpha ,\sin \alpha ,-a\cos \alpha )-
grw(\cos \alpha ,c\sin \alpha ,0),$$
$$\vert \vec{r}'\times \vec{r}''\vert^2 = r^4w^6+a^2r^4w^4\Omega^2\cos^2 \alpha +g^2r^2w^2(\cos^2\alpha +c^2\sin^2\alpha )-
2gr^3w^4a\cos \alpha +$$
$$+2acgr^3w^3\Omega \cos \alpha ,$$
and hence,

$$\vec{V}_b\approx -\lambda rw\vec{b}= $$
$$\frac{w^4r^4\sin\alpha [ar^3w^4\Omega (\Omega c-\frac{3}{2}w)-ga^2cr^2w^2\Omega^2\cos\alpha +
\frac{3}{2}g^2rw^2ac+3a^2r^2w^3\Omega g\cos\alpha ]}
{[r^4w^6+a^2r^4w^4\Omega^2\cos^2 \alpha +g^2r^2w^2(1-a^2\sin^2\alpha )
-2agr^3w^3\cos \alpha (w-c\Omega)]^2}$$
$$\cdot(r^2w^3\vec{b}^*-ar^2w^2\Omega\cos \alpha
(c\cos \alpha ,\sin \alpha ,-a\cos \alpha )-grw(\cos \alpha ,c\sin\alpha ,0))d\alpha .$$
Using the formula (\ref{gsv}) where $\vec{r}=(c\cos \alpha ,\sin \alpha ,-a\cos \alpha )$,
after some transformations,
the approximative spin velocity over the whole circle and for arbitrary $t$, can be written in the form

$$
\langle \vec{V}\rangle \approx -\frac{1}{\pi}\frac{r^2w^3}{g}(\vec{b}^*\times \vec{g}) $$
$$\cdot\int_{0}^{\pi}\frac{\sin^2\alpha [r^2w^2\Omega (\Omega c-\frac{3}{2}w)-gacr\Omega^2\cos\alpha
+\frac{3}{2}g^2c+3arw\Omega g\cos\alpha ](arw\Omega\cos \alpha +gc)d\alpha}
{[r^2w^4+a^2r^2w^2\Omega^2\cos^2 \alpha +g^2(c^2+a^2\cos^2\alpha )
-2agrw\cos \alpha (w-c\Omega)]^2}.$$
The spin motion of the spinning circle in this case is a circle with radius $R=\vert\langle \vec{V}_b\rangle \vert/\Omega$,
which can be tested.

(b) Apart from the gravitational acceleration we should apply also the acceleration caused by the horizontal plane which is
in the direction of the vector $\vec{b}^*$. It means we should add the acceleration $-\vec{b}^*(\vec{b}^*\cdot \vec{g})$,
i.e. if we decompose $\vec{g}$ in the direction of the axis and the other part, the direction of the axis should be neglected.
The new component should be replaced also in $\vec{r}'''$.
So, in this case we have

$$\vec{r}'=wr\vec{t},\qquad
\vec{r}''=-rw^2(\vec{t}\times \vec{b}^*) + \vec{t}\cdot r(\vec{w}'\cdot \vec{b}^*)-
\vec{b}^*(\vec{w}'\cdot \vec{t})r+\vec{g}-\vec{b}^*(\vec{b}^*\cdot \vec{g}),$$
$$\vec{r}'''=-[rw^3-r(\vec{w}''\cdot \vec{b}^*)]\vec{t}-3rw(\vec{w}'\cdot \vec{b}^*)(\vec{t}\times \vec{b}^*)+
\vec{b}^*(-(\vec{t}\cdot \vec{w}'')+\frac{3}{2}w[(\vec{t}\times \vec{b}^*)\cdot \vec{w}'])r-$$
$$-\frac{3}{2}w(\vec{g}\times \vec{b}^*)
-\frac{d\vec{b}^*}{dt}(\vec{g}\cdot \vec{b^*}).$$

Analogously as in case (a) we assume that $t=0$. The vectors
$\vec{b}^*$, $\vec{w}'$ and $\vec{w}''$, are the same as in case (a), while for the derivatives of $\vec{r}$ we obtain

$$\vec{r}'=wr(-c\sin \alpha ,\cos \alpha ,a\sin \alpha ),$$
$$\vec{r}''=-rw^2(c\cos \alpha ,\sin \alpha ,-a\cos \alpha ) -
arw\Omega \cos \alpha (a,0,c) -(0,0,g)+(a,0,c)gc,$$
$$\vec{r}'''=-(rw^3+ra^2w\Omega ^2)(-c\sin \alpha ,\cos \alpha ,a\sin \alpha )-
(a,0,c)ra\Omega w\sin \alpha (\Omega c-\frac{3}{2}w)+$$
$$+\frac{3}{2}awg(0,1,0)+gac\Omega (0,1,0).$$
Further, we obtain

$$\vec{r}'\times \vec{r}'''=
-ar^2w^2\Omega \sin \alpha (\Omega c-\frac{3}{2}w)(c\cos \alpha ,\sin \alpha ,-a\cos \alpha )-
\frac{3}{2}rw^2ag\sin \alpha \vec{b}^*-$$
$$-wrgac\Omega \sin \alpha \vec{b}^*,$$
$$(\vec{r}'\times \vec{r}''')\cdot \vec{r}''=
ar^3w^4\Omega \sin \alpha (\Omega c-\frac{3}{2}w)+\frac{3}{2}a^2r^2\Omega w^3g\cos \alpha \sin \alpha -$$
$$-ga^2r^2w^2\Omega \sin \alpha \cos\alpha (\Omega c-\frac{3}{2}w) +\frac{3}{2}g^2rw^2ac \sin \alpha +$$
$$+w^2r^2ga^2c\Omega^2\sin\alpha \cos\alpha -\frac{3}{2}rw^2ag^2c\sin \alpha ,$$
$$(\vec{r}',\vec{r}'',\vec{r}''')=-ar^2w^3\Omega\sin \alpha \bigl[ rw(\Omega c-\frac{3}{2}w)+3ag\cos\alpha \bigr] ,$$
$$\vec{r}'\times \vec{r}''=r^2w^3\vec{b}^*+wr(gc-arw\Omega\cos \alpha )(c\cos \alpha ,\sin \alpha ,-a\cos \alpha )-$$
$$-grw(\cos \alpha ,c\sin \alpha ,0),$$
$$\vert \vec{r}'\times \vec{r}''\vert^2 = r^4w^6+a^2r^4w^4\Omega^2\cos^2\alpha +g^2r^2w^2\cos^2\alpha a^2 -
2gr^3w^4a\cos \alpha +$$
$$+2acgr^3w^3(\Omega \cos \alpha -1)=r^2w^2N,$$
where

$$N=r^2w^4+a^2r^2w^2\Omega^2\cos^2\alpha +g^2\cos^2\alpha a^2 -
2grw^2a\cos \alpha + 2acgrw(\Omega \cos \alpha -1),$$
and hence

$$\vec{V}_b\approx -\lambda rw\vec{b}= \frac{a\sin \alpha r^2w^3\Omega \bigl[ rw(\Omega c-\frac{3}{2}w)+3ag\cos\alpha \bigr]}{N^2}$$
$$\cdot(r^2w^3\vec{b}^*+wr(gc-arw\Omega\cos \alpha )(c\cos \alpha ,\sin \alpha ,-a\cos \alpha )-
grw(\cos \alpha ,c\sin \alpha ,0)).$$
After some transformations, the approximate spin velocity over the whole circle and for arbitrary $t$, can be written
in the from

$$
\langle \vec{V}\rangle \approx -\frac{1}{\pi}\frac{r^4w^5a\Omega^2}{g}(\vec{b}^*\times \vec{g})
\int_{0}^{\pi}\frac{\sin^2\alpha \cos \alpha \bigl[ rw(\Omega c-\frac{3}{2}w)+3ag\cos\alpha \bigr] }{N^2}d\alpha .$$
The spin motion of the spinning circle in this case is a circle with radius $R=\vert\langle \vec{V}_b\rangle \vert/\Omega$,
which can be tested.

{\bf Remark 1.} We should remark about the following condition of equilibrium in case (b). While the body
moves with spin velocity $\vert \vec{V}\vert =V$, there appears centripetal acceleration $V\Omega$ toward the axis of rotation.
The resultant acceleration of this centripetal acceleration and the gravitational acceleration must have a direction collinear
with the vector $\vec{b}^*$. So, the angle $\varphi$ must satisfy the equilibrium condition (see Fig. 2)

\begin{figure}
\begin{center}
 \includegraphics[scale=0.7]{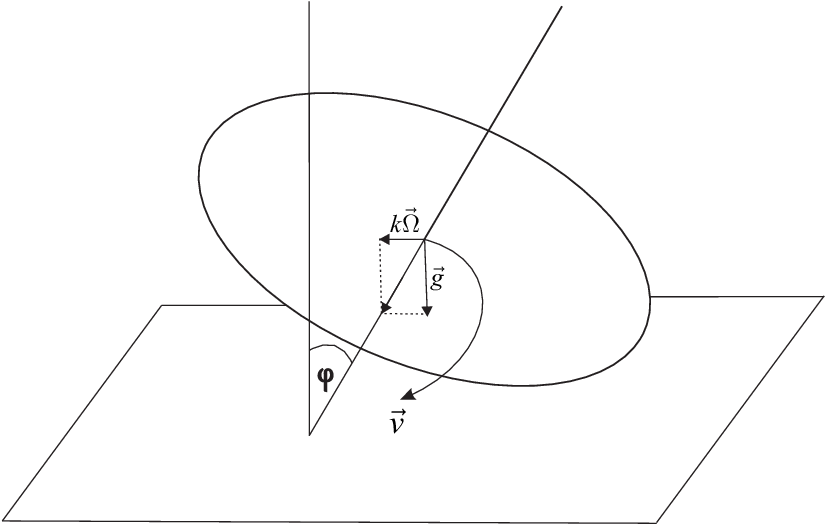}
 \label{fig:1}
 \caption{Condition of equlibrium: The resultant force of the gravitational force and the centripetal force should be directed
toward the spinning axis. }
\end{center}
\end{figure}

$$\tan \varphi =\frac{V\Omega}{g}.$$
When the body starts to rotate with small angle $\varphi$, then this angle $\varphi$ increases 
until the equilibrium condition is satisfied. If at the initial moment the angle 
$\varphi$ is such that $\tan \varphi >\frac{V\Omega}{g}$, then the spinning body can't continue and falls on the horizontal plane.

{\bf Remark 2.} The radius vector $\vec{r}$ in formula (\ref{gsv}), in general case, has the form $(r\cos \alpha ,r\sin \alpha ,h)$
in a coordinate system where at the chosen moment $\vec{b}^*=(0,0,1)$ and $h$ has the same meaning as in formula $(\ref{curve})$
and the initial point of the vector $\vec{r}$ is the barycenter (gravity center) of the spinning body. Since we considered
a spinning circle, the gravity center is at the center of the circle and $h=0$ in this case. Consequently, the spin vector
$\vec{V}$ was parallel to the horizontal plane. In general case, if someone considers a spatial body (not simply a planar body),
where the mass is not uniformly distributed along the spinning axis, then the vector $\vec{V}$ may also have a vertical component.
It may cause a situation where the gravity center of the body is on larger distance from the horizontal plane then its distance
at the initial moment (see Fig. 3). 

\begin{figure}
\begin{center}
 \includegraphics[scale=0.7]{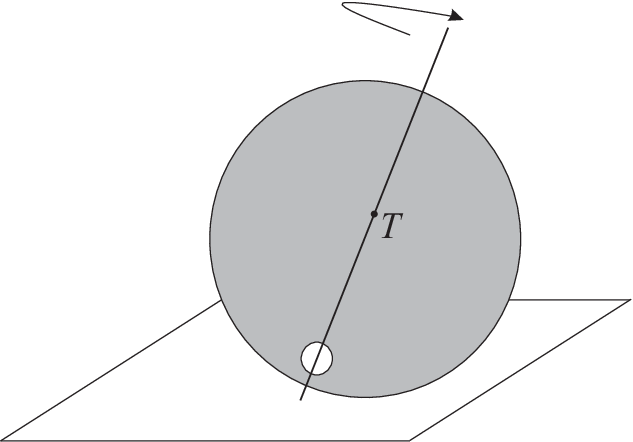}
 \label{fig:2}
 \caption{The gravity center is on larger distance of the horizontal plane opposite to our expectation.}
\end{center}
\end{figure}

\section{Spin velocity of the Earth}

The measurements of the length of day (LOD) show slight variations during a period of one year.
Using Fourier decomposition of the variation, there are two main components:
a) semiannual and b) annual variation. In case of semiannual variation of the LOD there is an amplitude of 0.29 ms maximizing on May 8,
while the annual variation of the LOD has amplitude of 0.34 ms, maximizing on February 3 (see for example \cite{Rosen}). It means that
in case of semiannual variation of the angular velocity of the Earth there is an amplitude of
$2.44\cdot 10^{-13}\;\textrm{s}^{-1}$ minimizing on May 8,
while the annual variation of the angular velocity of the Earth has an amplitude of $2.86\cdot 10^{-13}\;\textrm{s}^{-1}$ minimizing on February 3.

We will consider these two variations separately, because they are caused by different sources.

\subsection{Semiannual variation}

In case of semiannual variation the measurements show that the magnitude of the variation is zero in case of the
equinox and in case of summer and winter solstices. 
So, there is no doubt that this case is caused by the direction of the Earth's axis, as it is
tilted toward the ecliptic for the angle $\gamma =23.5^o$. We also use the following notations: $w$ is the angular velocity of the Earth
around its axis; $\Omega$ is the angular velocity of the Earth around the Sun, where in this subsection we may assume that $\Omega$
is a constant, i.e. the eccentricity is $\epsilon =0$; $R$ is the radius of the Earth, 
where we assume that the Earth is spherical body with constant density $\rho$; 
${\cal R}$ is the distance between the centers of the Sun and Earth; $r$ is the distance between
the center of the Earth and an arbitrary point from the interior of the Earth. The influences of the eccentricity $\epsilon$ and
non-homogeneity of the Earth's interior will be discussed later. 

Now, let us consider a spinning circle where $\vec{b}^*$ is a constant vector with coordinates $(0,0,1)$,
$\vec{w}=(0,0,w)=const$ and hence $\vec{w}'=\vec{w}''=0$. According to the formulas (\ref{r'}), (\ref{r''}) and (\ref{r'''})
we obtain

$$\vec{r}'=wr(-\sin \alpha ,\cos \alpha ,0),$$
$$\vec{r}''=-rw^2(\cos \alpha ,\sin \alpha ,0)+\vec{g},$$
$$\vec{r}'''=rw^3(\sin\alpha ,-\cos \alpha ,0)-\frac{3}{2}(\vec{g}\times \vec{w})+\frac{d\vec{g}}{dt}.$$

For each point inside the Earth, there are gravitational fields: toward the center of the Earth and toward the Sun. The acceleration
toward the center of the Earth is blocked, and so it should be skipped and only remains the gravitation with constant magnitude of
approximately $g=6\;\textrm{mm/s}^2$  changing only its direction. Since

\begin{equation}
\left [\begin{array}{ccc}\cos \gamma & 0 & -\sin \gamma \\
0 & 1 & 0\\
\sin \gamma & 0 & \cos \gamma \end{array}\right ]
\left [\begin{array}{ccc}\cos \Omega t \\ \sin \Omega t \\ 0 \end{array}\right ]
=\left [\begin{array}{ccc}\cos \Omega t \cos \gamma \\ \sin \Omega t \\ \cos \Omega t\sin \gamma \end{array}\right ] ,
\end{equation}
we may assume

$$\vec{g}=-(\cos \Omega t \cos \gamma ,\sin \Omega t,\cos \Omega t\sin \gamma )g.$$
Since $\Omega $ is about 365 times smaller than $w$, the term $\frac{d\vec{g}}{dt}$ may be neglected.
The coordinate system is chosen such that at equinoxes we have $\cos \Omega t=0$, or $\sin \Omega t=\pm 1$. 

Further, we obtain

$$\vec{r}'\times \vec{r}'''=\frac{3}{2}rw^2(-g_x\sin \alpha +g_y\cos \alpha )\vec{b}^*,$$
$$(\vec{r}',\vec{r}'',\vec{r}''')=\frac{3}{2}rw^2(g_x\sin \alpha -g_y\cos \alpha )(\vec{g}\cdot \vec{b}^*)=$$
$$=\frac{3}{2}rw^2g^2\cos \Omega t\sin \gamma (\cos \Omega t\cos\gamma \sin \alpha -\sin \Omega t\cos \alpha ),$$
$$\vec{r}'\times \vec{r}''=r^2w^3\vec{b}^*+wr(g_z\cos \alpha ,g_z\sin \alpha , -g_y\sin \alpha -g_x\cos \alpha )=$$
$$=wr(-g\cos \Omega t\sin \gamma \cos \alpha ,-g\cos \Omega t\sin \gamma \sin\alpha,$$
$$rw^2+g(\sin \alpha \sin \Omega t+\cos \alpha \cos \Omega t\cos \gamma )).$$

Let us introduce an angle $\alpha_0$ such that

$$\sin \alpha_0 = \frac{\sin\Omega t}{\sqrt{1-\cos^2\Omega t\sin^2\gamma}}, \qquad
\cos \alpha_0 = \frac{\cos\Omega t\cos \gamma}{\sqrt{1-\cos^2 \Omega t\sin^2\gamma}}.$$
Then,

$$\sin\alpha \cos \Omega t\cos \gamma -\cos \alpha \sin \Omega t
=\sqrt{\cos ^2\Omega t\cos ^2\gamma +\sin^2 \Omega t} \sin (\alpha -\alpha_0)=$$
$$=\sqrt{1-\cos^2 \Omega t\sin^2\gamma}\sin (\alpha -\alpha_0)$$
and

$$\cos \alpha \cos \Omega t\cos \gamma + \sin\alpha \sin \Omega t=
\sqrt{\cos ^2\Omega t\cos ^2\gamma +\sin^2 \Omega t} \cos (\alpha -\alpha_0)=$$
$$=\sqrt{1-\cos^2 \Omega t\sin^2\gamma}\cos (\alpha -\alpha_0).$$
Now, according to these notions, we obtain

$$(\vec{r}',\vec{r}'',\vec{r}''')=\frac{3}{2}rw^2g^2\cos \Omega t\sin \gamma
\sqrt{1-\cos^2 \Omega t\sin^2\gamma}\sin (\alpha -\alpha_0),$$
$$\vec{r}'\times \vec{r}'' = wr(-g\cos \Omega t\sin \gamma\cos \alpha ,-g\cos \Omega t\sin \gamma\sin \alpha ,$$
$$rw^2+g\sqrt{1-\cos^2 \Omega t\sin^2\gamma}\cos (\alpha -\alpha_0 )),$$
$$\vert \vec{r}'\times \vec{r}''\vert = wr\bigl[ g^2\cos^2 \Omega t\sin^2 \gamma +
(rw^2+g\sqrt{1-\cos^2 \Omega t\sin^2\gamma}\cos (\alpha -\alpha_0 ))^2\bigr ]^{1/2}.$$

If we replace $A=
\frac{3}{2}rw^2g^2\cos \Omega t\sin \gamma
\sqrt{1-\cos^2 \Omega t\sin^2\gamma}\sin (\alpha -\alpha_0)$ and $B=
g^2\cos^2\Omega t \sin^2 \gamma +
(rw^2+g\sqrt{1-\cos^2 \Omega t\sin^2\gamma}\cos (\alpha -\alpha_0 ))^2$, then the spin velocity
$\vec{V}_b=-rw\frac{k\tau}{k^2+\tau^2}\vec{b}$ takes the form

$$\vec{V}_b=-rw\frac{AB}{A^2+B^3}$$
$$\cdot(-g\cos \Omega t\sin \gamma\cos \alpha ,-g\cos \Omega t\sin \gamma\sin\alpha ,
rw^2+g\sqrt{1-\cos^2 \Omega t\sin^2\gamma}\cos (\alpha -\alpha_0 )).$$
If we divide by $g^6$ in the numerator and the denominator and use the replacement $X=\frac{rw^2}{g}$, the spin velocity takes the from

$$\vec{V}_b=-rw\frac{AB}{A^2+B^3}$$
$$\cdot(-\cos \Omega t\sin \gamma\cos \alpha ,-\cos \Omega t\sin \gamma\sin\alpha ,
X+\sqrt{1-\cos^2 \Omega t\sin^2\gamma}\cos (\alpha -\alpha_0 )),$$
where $A=\frac{3}{2}X\cos \Omega t\sin \gamma \sqrt{1-\cos^2\Omega t\sin^2\gamma}\sin (\alpha -\alpha_0)$ and
$B=\cos^2\Omega t\sin^2\gamma+(X+\sqrt{1-\cos^2\Omega t\sin^2\gamma}\cos (\alpha -\alpha_0))^2$.
The radius vector of an arbitrary point can be written as
$\vec{r}=(r\cos\alpha ,r\sin \alpha ,z)$ and in order to calculate the "global" spin velocity according to (\ref{gsv}) we find

$$\frac{(\vec{V}\cdot \vec{r})\vec{r}}{r^2}=rw\frac{AB}{A^2+B^3}\bigl[ r\cos \Omega t\sin \gamma -z(X+
\sqrt{1-\cos^2 \Omega t\sin^2\gamma}\cos (\alpha -\alpha_0)\bigr]$$
$$ \cdot\frac{(r\cos \alpha ,r\sin \alpha ,z)}{r^2+z^2}.$$
In order to average this term, we integrate by $\alpha\in [-\pi ,\pi ]$. Since $\alpha_0$ is a fixed angle, we replace
$\alpha - \alpha _0=\beta$, i.e. $\alpha =\alpha_0+\beta $ and integrate by $\beta\in [-\pi ,\pi ].$
Using that $\cos \alpha = \cos \beta\cos \alpha_0-\sin \beta\sin\alpha_0$ and
$\sin \alpha =\sin \beta\cos \alpha_0 + \sin\alpha_0\cos\beta$,
then for the averaging inside the Earth with arbitrary (not necessarily constant) density $\rho$, we obtain
$$
\langle \vec{V}\rangle = \Bigl[ \frac{1}{I}\int_{-R}^{R}dz \int_0^{\sqrt{R^2-z^2}}dr \int_{-\pi}^{\pi}d\beta \cdot
r\cdot rw\cdot \rho \frac{AB}{A^2+B^3}$$
\begin{equation}
\cdot\frac{r^2\cos \Omega t\sin\gamma-rz(X+\sqrt{1-\cos^2 \Omega t\sin^2\gamma}\cos \beta )}{r^2+z^2}\Bigr]
(-\sin\alpha_0 ,\cos \alpha_0,0),\label{density}
\end{equation}
where
$$I=\int_{-R}^{R}dz \int_0^{\sqrt{R^2-z^2}}dr \int_{-\pi}^{\pi}d\beta \cdot r\cdot \rho $$
is the mass of the Earth.
If $\rho$ is constant, then $I=\frac{4}{3}\pi R^3\rho$, and we obtain 
$$\langle \vec{V}\rangle = \Bigl[ \frac{1}{\frac{4}{3}\pi R^3}\int_{-R}^{R}dz \int_0^{\sqrt{R^2-z^2}}dr \int_{-\pi}^{\pi}d\beta \cdot
r\cdot rw\cdot \frac{AB}{A^2+B^3}$$
$$\cdot\frac{r^2\cos \Omega t\sin\gamma-rz(X+\sqrt{1-\cos^2 \Omega t\sin^2\gamma}\cos \beta )}{r^2+z^2}\Bigr]
(-\sin\alpha_0 ,\cos \alpha_0,0).$$

{\bf Remark 3.} It is known that he Earth interior is divided into 4 shells (\cite{earthint}):
inner core from 0 to 1200 km from the center,
outer core from 1200 to 3400 km,
stiffer mantle from 3400 to 5700 km,
and asthenosphere from 5700 to 6370 km.
Some thin shells close to the surface are omitted because they have no role with small volumes.
Later in order to calculate the spin velocity of the Earth,
will be used the following averaged densities:
inner core with density 13 g/cm$^3$,
outer core with density 11 g/cm$^3$,
stiffer mantle with density 5 g/cm$^3$, and
asthenosphere with density 3.4 g/cm$^3$. 

Analogously to $r=\frac{Xg}{w^2}$ we introduce also the replacement $z=\frac{Zg}{w^2}$ and using the Earth's constants as $R$, $w$, $g$
the averaging of the spin velocity becomes

$$\langle \vec{V}\rangle = \Bigl[ \frac{1}{\frac{4}{3}\pi 5.614^3}\frac{g}{w}\int_{-5.614}^{5.614}dZ
\int_0^{\sqrt{5.614^2-Z^2}}dX \int_{-\pi}^{\pi}d\beta \cdot
\frac{AB\sin\beta}{A^2+B^3}$$
$$\cdot\frac{X^3(X\cos \Omega t\sin\gamma -Z(X+\sqrt{1-\cos^2 \Omega t\sin^2\gamma}\cos \beta ))}{X^2+Z^2}\Bigr]
\frac{(-\sin\Omega t,\cos \Omega t\cos \gamma ,0)}{\sqrt{1-\cos^2 \Omega t\sin^2\gamma}}.$$
The term

$$Z(X+\sqrt{1-\cos^2 \Omega t\sin^2\gamma}\cos \beta ))$$
has no role in the integral and can be neglected. Using that

$$A=\frac{3}{2}X\cos \Omega t\sin \gamma
\sqrt{1-\cos^2 \Omega t\sin^2\gamma}\sin \beta$$
and after some arrangements, the averaging takes the form

$$\langle \vec{V}\rangle = \Bigl[ \frac{9\cos^2\Omega t\sin^2\gamma}{2\pi \cdot 5.614^3}\frac{g}{w}\int_{0}^{5.614}dZ
\int_0^{\sqrt{5.614^2-Z^2}}dX \int_{0}^{\pi}d\beta \cdot
\frac{B\sin^2\beta}{A^2+B^3}\frac{X^5}{X^2+Z^2}\Bigr]$$
$$\cdot(-\sin\Omega t,\cos \Omega t\cos \gamma ,0).$$

This vector can be written in the coordinate system in which the ecliptic plane coincides with the $xy$-plane, by multiplication
from left with the matrix

$$
\left [\begin{array}{ccc}\cos \gamma & 0 & \sin\gamma \\ 0 & 1 & 0 \\ -\sin\gamma & 0 & \cos \gamma \end{array}\right ].
$$
In the horizontal (ecliptic) plane the spin velocity is given by

$$\langle \vec{V}\rangle_h = \Bigl[ \frac{9\cos^2\Omega t\sin^2\gamma\cos \gamma}{2\pi \cdot 5.614^3}\frac{g}{w}\int_{0}^{5.614}dZ
\int_0^{\sqrt{5.614^2-Z^2}}dX \int_{0}^{\pi}
\frac{B\sin^2\beta}{A^2+B^3}\frac{X^5d\; \beta}{X^2+Z^2}\Bigr]$$
$$\cdot (-\sin\Omega t,\cos \Omega t ,0),$$
while the vertical component is given by

$$\langle \vec{V}\rangle_v = \Bigl[ \frac{9\cos^2\Omega t\sin^3\gamma}{2\pi \cdot 5.614^3}\frac{g}{w}\int_{0}^{5.614}dZ
\int_0^{\sqrt{5.614^2-Z^2}}dX \int_{0}^{\pi}
\frac{B\sin^2\beta}{A^2+B^3}\frac{X^5\; d\beta }{X^2+Z^2}\Bigr]$$
$$\cdot (0,0,\sin\Omega t).$$

We notice that the horizontal component is collinear with the ordinary velocity of motion of the Earth around the Sun. So,
{\em the spin velocity does not change the distance (at least for the first order of approximation) to the Sun}.
In case of $\epsilon =0$, the distance does not change. Only the
time parameter of the trajectory changes with the spin velocity.
For example, the spring or autumn equinoxes may occur earlier or later than expected.
The vertical component of the spin velocity shows that the Earth does not move in one plane.
Similar motion has the Sun in its trajectory in the Milky Way and the Moon on its trajectory around the Earth. 

We consider now the projection of the spin velocity on the ecliptic plane, i.e.
\begin{equation}
V_h =  \frac{9\cos^2\Omega t\sin^2\gamma\cos \gamma}{2\pi \cdot 5.614^3}\frac{g}{w}\int_{0}^{5.614}dZ
\int_0^{\sqrt{5.614^2-Z^2}}dX \int_{0}^{\pi}
\frac{B\sin^2\beta}{A^2+B^3}\frac{X^5\; d\beta}{X^2+Z^2}.\label{vh}
\end{equation}
It takes values between 0 (when $\cos \Omega t=0$, i.e. at the equinoxes) and takes its maximal value of order $1$ m/s
near the summer solstice and the winter solstice. This velocity may not be completely realized, 
but only partially, because of the influence from the Sun and the other planets.
We will explain this decreasing of the spin velocity according to the Sun and one planet, for example Jupiter, neglecting
temporarily the influence from the other planets.

Let $T_j$ be the gravity center between the Sun and Jupiter, let $d_j$ be the distance between
$T_j$ and the center of the Sun and let $m_j$ be the mass of Jupiter.
Let us denote by $U$ the spin velocity of the Earth, having in mind the influence of the solar system.
So, the
angular velocity of the Earth will be $U/{\cal R}$. Because of the gravitation between the Earth and the Sun, the distance Earth-Sun
should be preserved and so, the Sun will be rotated with the same angular velocity $U/{\cal R}$ around the point $T_j$.
Jupiter will not be rotated as the Sun, because the gravitation force between the Earth and Jupiter is too small to preserve the
distance Earth-Jupiter. So, Jupiter will be only translated as the Sun, such that the direction Sun-Jupiter will change only
according to their mutual gravitational force. We denote by $I$ the moment of inertia of the Sun around the axis of rotation
through the point $T_j$. Using that the distance between the Earth and $T_j$ is close to ${\cal R}$, we obtain the following equation

$$mV_h{\cal R}=mU{\cal R}+\frac{U}{{\cal R}}(I+Md_j^2),$$
and using that $d_j=\frac{m_jr_j}{M+m_j}\approx \frac{m_jr_j}{M}$, we obtain that $U=\frac{V_h}{k}$, where $k$ is the following
coefficient

$$k=1+\frac{I}{m{\cal R}^2} +\frac{m_j^2r_j^2}{Mm{\cal R}^2}$$
Now, we conclude that $\frac{I}{m{\cal R}^2}$ is the influence of the Sun,
while $\frac{m_j^2r_j^2}{Mm{\cal R}^2}$ is the influence of Jupiter in decreasing the spin velocity
$V_h$. The distance $r_j$ and the mass $m_j$ are well known and if we take that 
$I=5.7\cdot 10^{46}$ kg$\cdot$m$^2$ according to the available data (for example \cite{SunI}),
we obtain that $k=1+0.426+8.22=9.646$. We notice that the influence from Jupiter 
is much larger than the influence from the Sun. Indeed Jupiter has its influence through 
the change of the barycenter. Its influence through the acceleration will be
discussed later with the discussion of the influence of the Moon. 

Now, let us consider the influence of the other planets. Analogously to the coefficient 
8.22 for Jupiter, the corresponding coefficients for Saturn, Neptune and Uranus 
are $2.466$, $0.7934$ and $0.229$ respectively, while the influence from the other planets are 
negligible. Hence we can conclude that the total influence of the Solar system is approximately
$k=1+0.426+8.22+2.466+0.7934+0.229=13.138$,
i.e. the real spin velocity of the Earth is $U=\frac{V_h}{13.14}$. Now we should explain why we consider the influences from the
planets mutually independent. Indeed, one can suggest that instead of the gravity center between the Sun and Jupiter
to consider the gravity center between the Sun and all planets except the Earth. This is not convenient because the planets
do not preserve the distances between each pair of them and
so, they participate independently in decreasing of the spin velocity $V_h$.
In case of the Moon it should not have analogous contribution to the planets because it is more related directly to the
Earth via a binary system, than through the barycenter.

We will prove that the spin velocity induces angular velocity in the ecliptic plane which is given by
\begin{equation}
W=-\frac{U'}{2v},\label{W}
\end{equation}
and the direction is also orthogonal to the ecliptic plane, where $v=30$ km/s is the velocity of the Earth in the ecliptic plane.
Assuming that $v$ is constant, we make deviation of about $\pm \epsilon /2=0.83\%$.
The obtained angular velocity $W$ should be multiplied by $\cos \gamma = 0.917$,
because the observed length of the day is measured
with respect to the spinning Earth, but not with respect to the ecliptic plane.

The required formula will be obtained via the law of preserving the energy. There are three energies of the Earth:
$$E_1=m_e\vert (\vec{{\cal R}}\times (\vec{v}+\vec {U}))'\vert $$
where $\vec{v}$ is the velocity of 30 km/s
around the Sun. Since $\vec{{\cal R}}\times \vec{v}$ is a constant, only the spin velocity has influence, i.e.
$$E_1=m_e\vert (\vec{{\cal R}}\times \vec {U})'\vert =m_e{\cal R}U'.$$
This (potential) energy appears in case of perturbations in the classical Newtonian orbits in gravitational field.
The second energy is the classical one
$$E_2=m_e(\frac{1}{2}v^2-\frac{GM_s}{{\cal R}}).$$
Since we consider circular trajectory, it is convenient to write in the form
$$E_2=m_e(\frac{1}{2}v^2-{\cal R}^2\Omega ^2).$$

This energy remains unchanged for a classical trajectories without perturbations.
In case of additional angular velocity $W$ locally of the Earth, not necessary on the trajectory, the Earth will move with the
same inertial velocity $v$ and the angular velocity will be $\Omega +W$. It leads to the third energy
$$E_3=m_e\Bigl [\frac{1}{2}v^2-{\cal R}^2(\Omega +W)^2\Bigr ].$$
When appears the spin velocity $U$, instead of the energy $E_2$ we have the energy 
$E_3$ and the law of preserving the energies states that $E_3=E_2+E_1$, i.e.
$$m_e\Bigl [\frac{1}{2}v^2-{\cal R}^2(\Omega +W)^2\Bigr ]=m_e(\frac{1}{2}v^2-{\cal R}^2\Omega ^2)+m_e{\cal R}U'.$$
Since $W$ is very small, $W^2$ may be neglected and this equation can easily be written in the form
$$\frac{U'}{{\cal R}}+2\Omega W =0,$$
$$W= -\frac{U'}{2{\cal R}\Omega}=-\frac{U'}{2v}.$$

Further we discuss the influence of the Moon to the spin velocity of Earth.
According to the astronomical data, it is easy to calculate that the acceleration of the Earth toward the Moon is about 187 times
smaller than the acceleration $g$ toward the Sun. It changes mainly the magnitude of $g$, because the direction has a minor role.
According to (\ref{vh}) and assuming that the function under the triple integral is almost constant as a function of $g$, we conclude
that the spin velocity of the Earth should be approximately 187 times smaller, which gives variation of the magnitude of about
$0.53\%$, i.e. its corresponds in variation of 0.15 ms with a period of one month. The acceleration of the Earth toward the Jupiter,
or another planet, is at least 90 times smaller than the acceleration toward the Moon. So the planets do not have
any additional influence than that via the solar barycenter.
Beside the annual and semiannual variations in length of the day, it is also measured
variation of period of 10 days and maximal amplitude of about 0.1 ms. 

In Fig.  4 is given the graph of the function
$W$ as a function of $\alpha =\Omega t$, where $\Omega t=\pi /2$ corresponds to the spring or autumn equinox.
The calculation is done using the data about the densities of the shells from Remark 3, where the integration is analogous to
(\ref{density}). The amplitude of this
function $W$ is $2.516\cdot 10^{-13}\;\textrm{s}^{-1}$. Further this values should be multiply with
$\cos 23.5^0$ because the variation of the length of the day is measured from the Earth. Hence we obtain for the amplitude
$2.307\cdot 10^{-13}\;\textrm{s}^{-1}$, which is $94.56\%$ of the value
$2.44\cdot 10^{-13}\;\textrm{s}^{-1}$, which is indirectly measured via the length of day. Hence the predicted value is
$5.44\%$ less that the measured value. The maximal value of the LOD achieves at May 8 also fits with the observations.
We commented that the main reason for departures as the influence of Moon and the eccentricity are less than $1\%$.
Also the precise knowledge of the moment of inertia of the Sun will improve the expected value.
The rest influence is probably caused by the atmosphere.
But we can not do anything with the influence of the Kuiper belt. The atmospheric influence is not negligible.  
\begin{figure}
\begin{center}
 \includegraphics[scale=0.7]{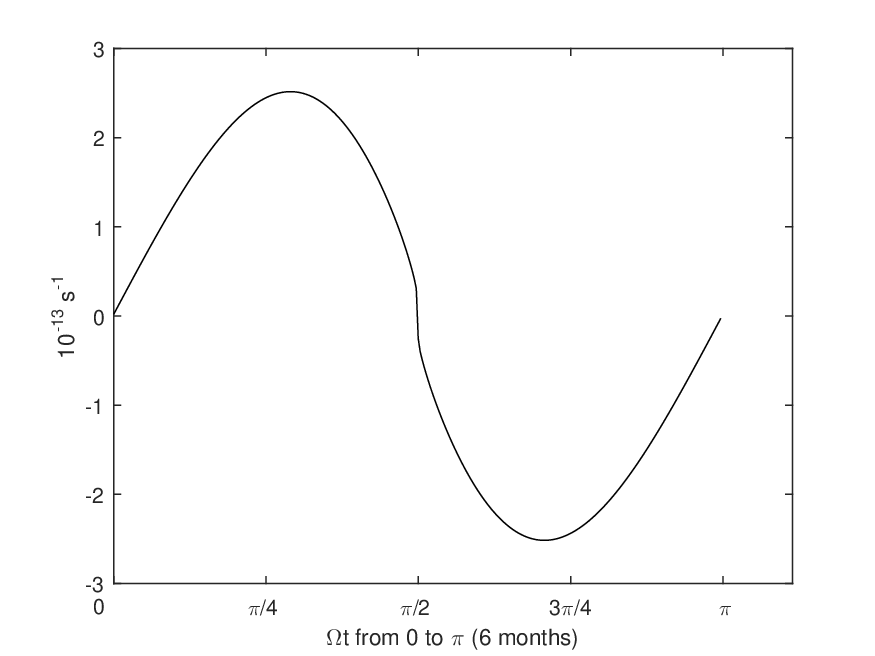}
 \label{fig:1}
 \caption{Graph of the function $W$ as a function of $\alpha =\Omega t$.
The length of day achieves its maximum at angle $3\pi /4$, i.e. about 45 days after each equinox ($\varphi =\pi /2$),
which means that the angular velocity achieves its minimal value. }
\end{center}
\end{figure}

{\bf Remark 4.} In the previous calculations we assumed that the Earth is a rigid body. 
The existence of the liquids inside the Earth and the surface water from the oceans and the seas changes the previous results.
In general when we calculate the resultant spin velocity, many components
in case of rigid body are mutually canceling. But in case of the surface liquids, it tends
to move and hence there appear some turbulence.
Probably this is related with the effect of tides, but this is only a suggestion for further research. 

\subsection{Annual variation}

This variation is of different nature, and it is a relativistic effect,
including the velocity of the light.
We refer to the papers \cite{Ukraina,India}, however, we start with the necessary preliminaries.
It is convenient to imagine the
Earth as a cube or square with edge $\Delta {\cal R}$ in the ecliptic plane (Fig. 5a).
Indeed, the final result will not depend on the shape and the size of the considered body. 
The largest and the smallest distances to the Sun are ${\cal R}+\Delta {\cal R}$ and ${\cal R}$ respectively.
Since the Earth is moving with an almost constant velocity $v$, then instead the orbits with lengths
$2\pi ({\cal R}+\Delta {\cal R})$ and $2\pi{\cal R}$ the points $B$ and
$A$ pass orbits with lengths $2\pi (r+\Delta r)$ and $2\pi r$ such that (Fig. 5b)

\begin{figure}
\begin{center}
 \includegraphics[scale=0.7]{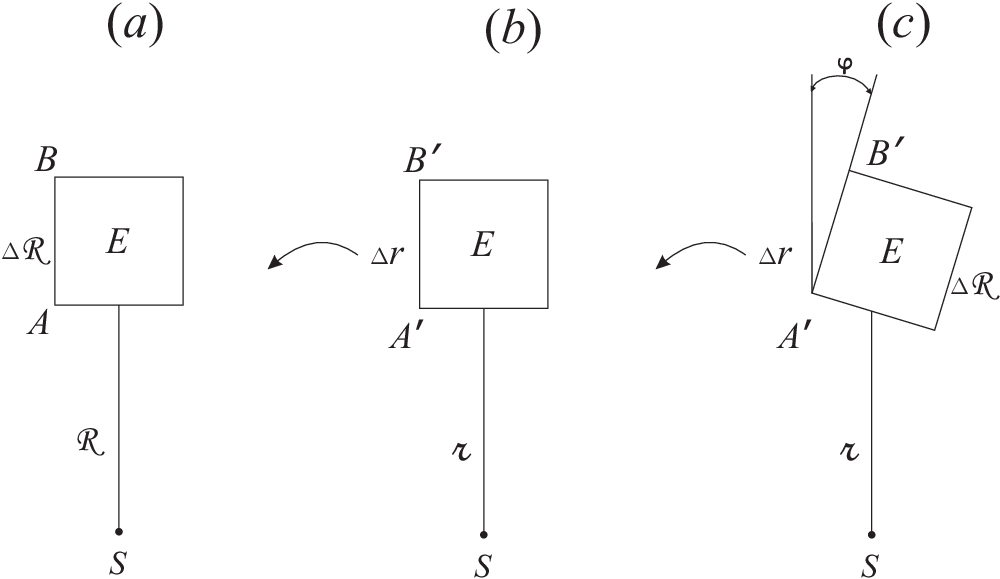}
 \label{fig:4}
 \caption{Position of the Earth toward the Sun. In case (a) it is assumed that the Earth is not moving.
In case of (b) the distance ${\cal R}$ changes into $r$ such that the contraction of the length of the trajectory in motion
$2\pi {\cal R}$
seems to be observed as a trajectory with smaller radius $r$. This can easily be evident in case of a spinning rotational disc.
In case (c) there should appear a rotation for angle $\varphi$, since according to the Special Relativity there is no contraction
orthogonal to the motion. If there is no rotation, there will be a contradiction to this relativistic effect.}
\end{center}
\end{figure}

\begin{equation}
{\cal R}\sqrt{1-\frac{v^2}{c^2}}=r,\label{1}
\end{equation}
where $v=rw$ and hence,
\begin{equation}
r=\frac{{\cal R}}{\sqrt{1+\frac{{\cal R}^2w^2}{c^2}}}. \label{2}
\end{equation}
Since there does not exist contraction orthogonal to the direction of motion $SA'$, then there will exist locally a rotation
for angle $\varphi$ (Fig. 5c) such that
$\cos \varphi =\frac{r}{\cal {R}}$ and
$\varphi <0$. Indeed, the point $B'$ appears later in the time, after the appearance of $A'$. Namely, there is a time delay of $B'$
with respect to $A'$.

The motion of the Earth in the solar system satisfies $\frac{v^2}{2}-\frac{GM}{{\cal R}}=C=const$ and hence,
\begin{equation}
v=\sqrt{2C+\frac{2GM}{{\cal R}}}.\label{3}
\end{equation}
Since the trajectory is close to circle, we can suppose that ${\cal R}w=v$, and from (\ref{2})

$$\frac{r}{\cal {R}}\approx 1-\frac{1}{2}\frac{v^2}{c^2}=\cos \varphi $$
and hence, $\varphi ^2=\frac{v^2}{c^2}$ and $\varphi \approx -\frac{v}{c}$, since $\varphi <0$.
Thus, for the angular velocity from (\ref{3}) we obtain

$$W=\frac{d\varphi}{dt}=-\frac{1}{c}\frac{dv}{dt}=\frac{1}{c}\frac{1}{2v}\frac{2GM}{{\cal R}^2}\frac{d{\cal R}}{dt}.$$
Using that the eccentricity $\epsilon \approx 0.0167$ is small, approximately it holds
${\cal R}={\cal R}_0(1-\epsilon \cos \Omega t)$, where ${\cal R}_0$ is a constant and $\Omega t =0$ on January 3-4,
when the Earth is on the closest distance to the Sun. So,

$$W=\frac{1}{cv}\frac{GM}{{\cal R}^2}{\cal R}_0(-\epsilon \Omega )\sin\Omega t = -\frac{g\epsilon}{c}\sin \Omega t,$$
and hence $\vert W\vert_{max}=\frac{g\epsilon}{c}$. Since this angular velocity is in the ecliptic plane, the predicted value is

$$\vert W\vert_{max}=\frac{g\epsilon}{c}\cos 23^0.5 = 3.022\cdot 10^{-13}\;\textrm{s}^{-1},$$
while the maximal amplitude is $5.36\%$ less. Probably the departure appears from the atmosphere.

Now we will determine the phase displacement of the obtained sinusoidal function. For this reason we will calculate
approximately the vector of velocity $\vec{v}$, apart from the determination of its module.

Starting from 
$$\vec{\cal R}=
{\cal R}_0\Omega (\cos \Omega t,\sin \Omega t,0)+{\cal R}_0(1-\epsilon\cos \Omega t)\Omega(-\sin \Omega t, \cos \Omega t,0),$$
we obtain

$$\vec{v}=\frac{d{\cal R}}{dt}={\cal R}_0\epsilon \Omega \sin \Omega t(\cos \Omega t,\sin \Omega t,0)
+{\cal R}_0(1-\epsilon \Omega t)\Omega (-\sin \Omega t,\cos \Omega t,0).$$

Now we make a distinction between $v_{tan}$, which is the coefficient in front of $(-\sin \Omega t,\cos \Omega t,0)$
and $v_{rad}$, which is the coefficient in front of $(\cos \Omega t, \sin \Omega t,0)$. In case of tangent component
we have $\varphi_{tan}=-\frac{1}{c}v_{tan}$, but in case of radial component ($v\neq const$ from
$r=\frac{{\cal R}}{\sqrt{1+\frac{{\cal R}^2w^2}{c^2}}}$ we obtain $dr=(d{\cal R})\bigl( 1-\frac{3}{2}\frac{({\cal R}w)^2}{c^2}\bigr)$,
and thus, from $\frac{dr}{d{\cal R}}=\cos \varphi$ it follows $\varphi_{rad}^2=3(\frac{{\cal R}w}{c})^2$,
$\varphi_{rad}=\sqrt{3}\frac{v}{c}$. So, in both cases we introduce a coefficient of proportionality $k$, such that

$$\varphi'_{tan}=-\frac{k}{c}\frac{dv_{tan}}{dt}=-\frac{k}{c}\frac{d[{\cal R}_0\Omega (1-\epsilon\cos \Omega t)]}{dt}=$$
$$\frac{k}{c}{\cal R}_0\Omega^2\epsilon (-\sin \Omega t)=k\frac{g\epsilon}{c}(-\sin \Omega t)$$
and

$$\varphi'_{rad}=-\frac{k\sqrt{3}}{c}\frac{dv_{rad}}{dt}=-\frac{k\sqrt{3}}{c}\frac{d[{\cal R}_0\Omega \epsilon \sin \Omega t)]}{dt}=
$$
$$
-\frac{k\sqrt{3}}{c}{\cal R}_0\Omega^2\epsilon (\cos \Omega t)=k\frac{\sqrt{3}g\epsilon}{c}(-\cos \Omega t).$$
The coefficient $k>0$ should be determined from the condition

$$\max (\varphi ') =\max (\varphi '_{tan}+\varphi '_{rad}),$$
i.e. $\varphi'$ and $\varphi '_{tan}+\varphi '_{rad}$ differ only by the phase. So,

$$\frac{g\epsilon}{c}=\max(k\frac{g\epsilon}{c}\sin\Omega t + k\frac{3g\epsilon}{c}\cos \Omega t)$$
$$1=k\cdot \max(\sin\Omega t + \sqrt{3}\cos\Omega t)=2k\cdot \max(\sin (\Omega t+\frac{\pi}{3})=2k.$$
So, $k=\frac{1}{2}$ and

$$W=-\frac{kg\epsilon}{c}(\sin \Omega t+\sqrt{3}\cos \Omega t) =-\frac{g\epsilon}{c}
(\frac{1}{2}\sin\Omega t + \frac{\sqrt{3}}{2}\cos\Omega t)=-\frac{g\epsilon}{c}\sin (\Omega t+\frac{\pi}{3}),$$
and its projection on the spinning axis is given by

$$W=-\frac{g\epsilon}{c}\sin (\Omega t+\frac{\pi}{3})\cos \gamma .$$
So, $W$ has a period $T=\frac{2\pi}{\Omega }=1$y, and it minimizes for $\Omega t=\frac{\pi}{6}$, 
starting from the January 3 when the Earth is the closest to the Sun and then $\Omega t=0$. 
Thus the minimum happens after $\frac{365.25}{12}\approx 30.4$ days
when the Earth will be the closest to the Sun, and it is about February 3.

\section{Conclusion}
This paper has mainly two objectives. The first objective is to present the recently 
developed theory about the so called induced spin velocity in the frame of a multidimensional 
space-time. The second objective is to apply this spin velocity in some concrete cases. 
As a consequence it is a test for the theory. Firstly, it is applied on a spinning body in a 
gravitational field where there is a precession of the spinning axis for both cases: 
when the spinning body is in free fall motion and when the spinning body is placed 
on a horizontal plane. The next application is explanation of the semiannual variation 
of the length of day. Although there are several perturbations of this effect, it is explained 
with about $5\%$ departure of observational data and it is the first comparison of the spin 
velocity with some experimental data. In the paper it is considered also the annual variation 
of the length of day, although it is not a consequence of the spin velocity. 
It is a consequence of the results in the recently published papers in order to avoid the 
known question whether the boundary points on a spinning disc with large radius will move 
with velocity larger that $c$. In this case the departure between the predictions and measurements is also about $5\%$.

\end{document}